# Volatility and Returns in Korean Futures Exchange Markets

Kyungsik Kim*, †, Seong-Min Yoon† and Jum Soo Choi†

†*Department of Physics, Pukyong National University, Pusan 608-737, Korea*

*Division of Economics, Pukyong National University, Pusan 608-737, Korea*

## ABSTRACT

We apply the formalism of the continuous time random walk (CTRW) theory to financial tick data of the bond futures transacted in Korean Futures Exchange (KOFEX) market. For our case, the tick dynamical behaviors of the returns and volatility for bond futures are treated particularly at the long-time limit. The volatility for the price of our bond futures shows a power-law with anomalous scaling exponent, similar to other options. Our result presented will be compared with that of recent numerical calculations.

*Corresponding author. Tel: +82-51-620-6354; fax: +82-51-611-6357.
E-mail address: kskim@pknu.ac.kr.

Recently, the investigation of the continuous time random walk (CTRW) theory has received considerable attention as one interdisciplinary field between economists and physicists [1,2]. This subject can mainly lead to a better understanding for scaling properties based on novel statistical methods and approaches of economics. Particularly, in the stochastic process, one of the well-known problems for the regular and disordered systems is the random walk theory [3], and this theory has extensively developed to CTRW theory, formerly introduced by Montroll and Weiss [4], which is essentially described both by the transition probability dependent of the length between steps and by the distribution of the pausing times [5]. Until now, CTRW theory has extensively studied in natural and social sciences, and among many outstanding topics, there has been mainly concentrated on the reaction kinetics [6] and the strange kinetics [7], fractional diffusion equations [8], random networks, earthquake modeling, hydrology, and financial options [9].

Recently, Scalas [10] presented the correlation function for bond walks from the time series of bond and BTP (Buoni del Tesoro Poliennali) futures exchanged at the London International Financial Futures and Options Exchange (LIFFE). Scalas *et al*. [11] have discussed that CTRW theory is applied to the dynamical behavior by tick-by-tick data in financial markets. Mainardi *et al*. [12] have observed the waiting-time distribution for bond futures traded at LIFFE, London. Kim and Yoon [13] studied the dynamical behavior of tick volume data for the bond futures in the Korean Futures Exchange market　(KOFEX) and obtained that the decay distributions for the survival probability display novel stretched-exponential forms. Very recently, Scalas *et al*. [14] applied CTRW theory to models of the high-frequency price dynamics and showed that the waiting-time survival probability for high-frequency data of 30DJIA stocks is non-exponential.

In this paper, theoretical and numerical arguments for the bond futures traded at KOFEX are presented, and we apply the formalism of CTRW theory to financial tick data of the bond futures. Our underlying asset is a Korean Government bond futures traded for KTB212, which were taken from July 2002 to December 2002 at the KOFEX. Particularly, the tick dynamical behavior of the volatility and returns for the bond futures are treated at long-time limits.

First of all, we will focus on CTRW theory in order to discuss the probability density function. Let $R(t)$ be the return defined by

$$R(t) = \ln[P(t+t_0)/P(t_0)]$$ 　　　　（ 1 ）

where $P(t)$ is the price of KTB212 at time $t$, and $t_0$ is an arbitrary time. The zero-mean return $Y(t)$ becomes

$$Y(t) = R(t) - <R(t)> \qquad (\,2\,)$$

In our case, we introduce the pausing time density function and the transition probability. When $X_n = t_n - t_{n-1}$ and $\Delta Y_n = Y(t_n) - X(t_{n-1})$, the pausing time density function and the jumping probability density function are, respectively, defined by

$$\psi(t)dt = \mathrm{Pr}\,ob\,\{\,t < X_n < t+dt\,\} \qquad (\,3\,)$$

$$j(l)dl = \mathrm{Pr}\,ob\,\{\,l < \Delta Y_n < l+dl\,\}\;. \qquad (\,4\,)$$

We also consider that the coupled probability density $\Psi(l, t)$ for jump $l$ and pausing time $t$ is defined by

$$\Psi(l, t) = j(l)\,\psi(t)\;, \qquad (\,5\,)$$

where $j(l)$ is the jumping probability density dependent of the length between steps and $\psi(t)$ is the pausing-time density. Summing over all $l$ with the periodic boundary condition, Eq. (1) is given by

$$\sum_l \Psi(l, t) = \sum_l j(l)\,\psi(t) = \psi(t)\;. \qquad (\,6\,)$$

Since $R_n(l, t)$ is the probability density arriving immediately at lattice point $l$ at time $t$ after one random walker goes to $n$ steps, $R_n(l, t)$ is satisfied with the recursion relation as

$$R_n(l, t) = \sum_l \int_0^t dt'\, j\,(l-l')\,\psi\,(t-t')\,R_n(l, t) \qquad (\,7\,)$$

In order to find the probability density function (PDF) $P(l, t)$, which is existed at lattice point $l$ at time $t$, we introduce the relation between $P(l, n)$ and $R(l, t-t')$ such as

$$R(l, t) = \sum_{n=0}^{\infty} \psi_n(t)\,P(l, n)\;. \qquad (\,8\,)$$

The survival probability $\Psi(t)$, which is stayed for the time $t$ after arriving at an arbitrary lattice point, can be expressed in terms of

$$\Psi(t) = 1 - \int_0^\infty dt \sum_l \psi(l, t) = 1 - \int_0^\infty dt \, \psi(t) \ . \qquad (\,9\,)$$

Hence, after substituting Eq. (6) into Eq. (5), the Fourier-Laplace transform of the PDF, i.e., Montroll-Weiss equation, is described [16] as

$$P(k, u) = \sum_l \int_0^\infty dt \, P(l, t) \, \exp(-ikl - ut)$$
$$= P(k, 0)[1 - \psi(u)] / \, u[1 - \Psi(k, u)] \ , \qquad (\,10\,)$$

where $P(k, 0)$ is the Fourier transform of the initial condition $P(l, 0)$, and the generalized structure function, i.e., the Fourier-Laplace transform of Eq. (1) is given by

$$\Psi(k, u) = j(k) \, \psi(u) \qquad (\,11\,)$$

For example, Let us introduce the jumping probability density

$$j(l) = p \, \delta(l+1) + q \, \delta(l-1) \qquad (\,12\,)$$

and two pausing time densities given by

$$\psi_1(t) = 2a \, \pi^{-1/2} \, \exp(-a^2 t^2) \qquad (\,13\,)$$

and

$$\psi_2(t) = 4a^2 \, \exp(a^2 t) \, i^2 \, erfc \, (a t^{1/2}) \ . \qquad (\,14\,)$$

The PDFs of Eq. (10), through the inverse Fourier-Laplace transform of Eq. (10), can be found as

$$P_1(l, t) \cong (2\pi \overline{l^2} t / \overline{t})^{-1/2} \, \exp(-l^2 / 2 \overline{l^2} (t / \overline{t})) \qquad (\,15\,)$$

and

$$P_2(l, t) \cong (2\pi a^2 \overline{l^2} t) \, \exp(-l^2 / 2 a^2 \overline{l^2} t)$$

$$+ (2/\overline{l^2})^{1/2} \, erfc(l/(2a^2\overline{l^2}t)^{1/2}) \, . \qquad (16)$$

Here $\overline{l^2}$ is the second moment of the jumping probability density, and $\overline{t}$ is the first moment of the pausing time density, and $erfc\,(x) = 1 - erf\,(x)$, where $erf(x)$ is the error function.

In our model, if the jumping probability density and the pausing time density in the high-frequency region are, respcetively, given by the power-law such as

$$j\,(l) \sim l^{-\beta} \qquad (17)$$

and

$$\psi\,(t) \sim t^{-\alpha-1} \, , \qquad (18)$$

at the long time limit, then the PDF becomes

$$P\,(l,\,t) \sim |l|^{-\beta} \qquad (19)$$

as $|l| \to \infty$. The volatility, i.e., the return variance, is derived as

$$<l^2(t)> \sim t^{\alpha} \, . \qquad (20)$$

Hence we can analyze numerically the dynamical behavior of the PDF and the volatility from our tick data, based on the CTRW theory.

In order to analyze the scaling exponents of the statistical quantities, we apply the formalism of the CTRW theory to the KTB112. Fig.1 shows the continuous tick data of volumes for KTB212 for six months, which were taken from July 2002 to December 2002. We numerically find that the pausing time density of KTB212 is proportional to a power law $\psi(t) \sim t^{-\alpha-1}$ with the scaling exponent $\alpha = 0.51$. The jumping probability density is given by a power law $h(l) \sim l^{-\beta}$ with $\beta = 2.94$ from the continuous tick data of the price for KTB212, as plotted in Fig.3.

For the PDF, we obtained the scaling exponents $\beta = 3.45$, 3.68, 3.10, and 2.41 for one time lag $t = 1$ second, 1 minute, 10 minutes and 1 hour. Interestingly, the return variation shows the power-law with the scaling exponent $\alpha = 0.34$ in Fig.6, and the scaling exponent of the pausing time density obtained from our tick data is not exactly the same value as that of the volatility. Moreover, our volatility has a subdiffusive behavior, while the result of Masoliver *et al* [9] has a diffusive-like behavior at the long time limit.

In conclusion, we have presented the dynamical behavior of the high-frequency tick data for the KTB212, based on a continuous-time random walk theory. We find numerically the pausing time density and the jumping probability density. Particularly, the PDF and the return variation scale as a power law with scaling exponents for large times. It is expected in future that the detail description of the CTRW will be used to study the extension of financial analysis in Korean financial markets.

# Figure Captions

**Fig.1**   The continuous tick data of volumes for KTB212 were taken from July 2002 to December 2002.

**Fig.2**   Pausing time density of KTB212 for one time step of 1 second is given by a power law $\psi(t) \sim t^{-\alpha-1}$ with $\alpha = 0.51$.

**Fig.3**   The continuous tick data of the price $P(t)$ for one time step of 10 minutes were taken from July 2002 to December 2002.

**Fig.4**   Jumping probability density for KTB112 is given by a power law $j(l) \sim l^{-\beta}$ with $\beta = 2.94$.

**Fig.5**   PDFs $P(l,t)$ of KTB112 for a time lag $t = 1$ second, 1 minute, 10 minutes and 1 hour are proportional to a power law with $\beta = 3.45$, 3.68, 3.10, and 2.41.

**Fig.6**   Return variance for KTB112 is proportional to a power law with $\alpha = 0.34$ .

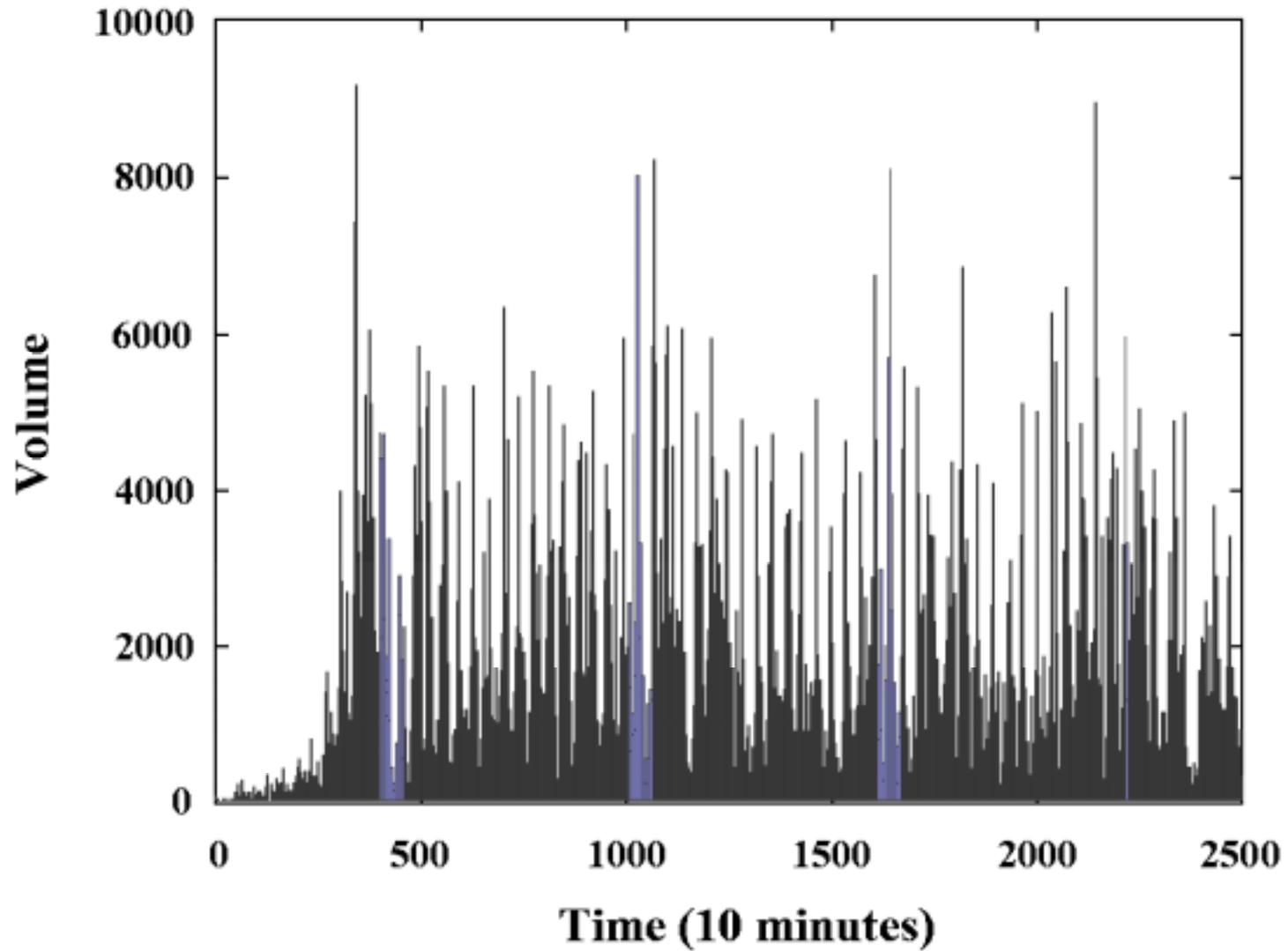

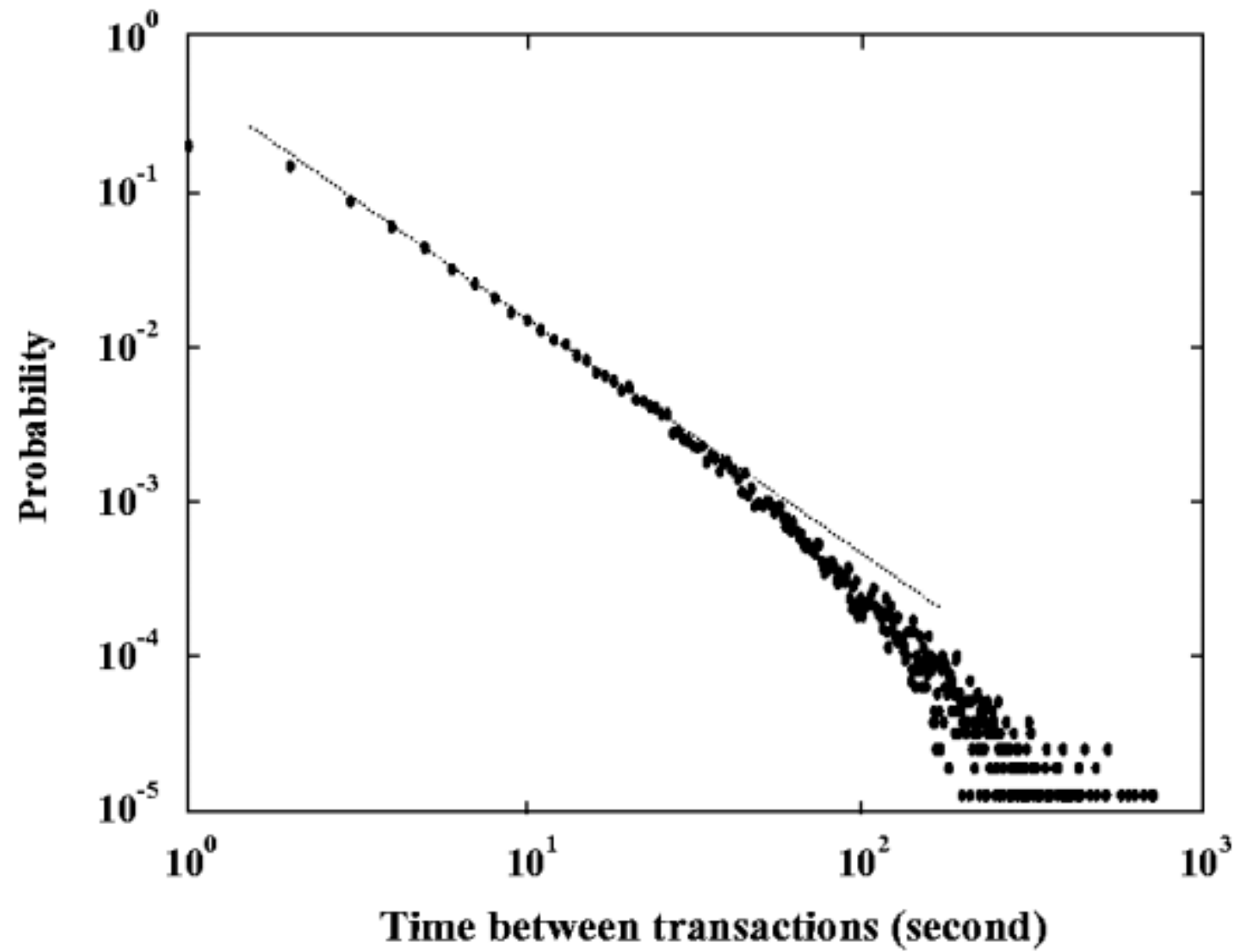

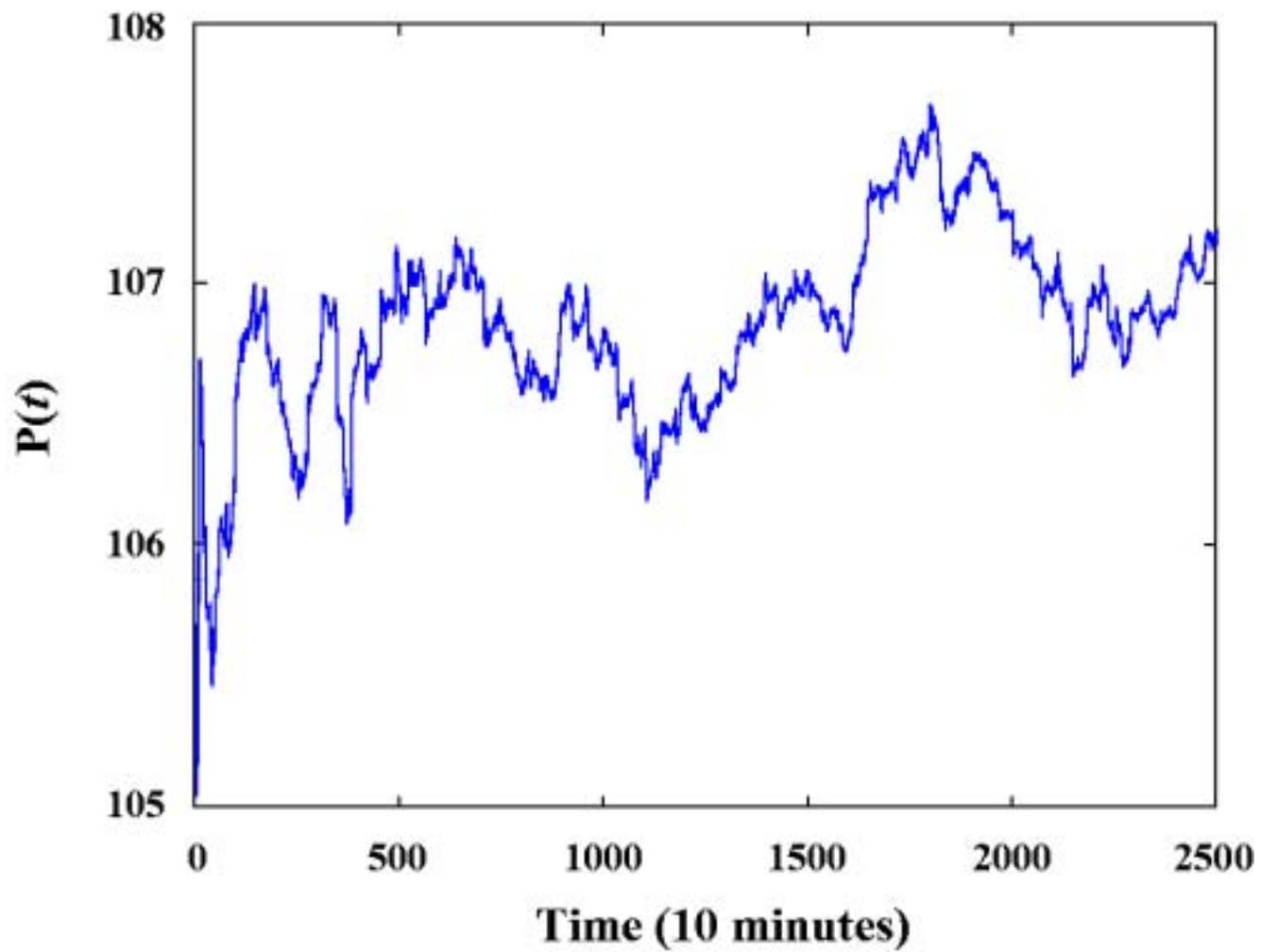

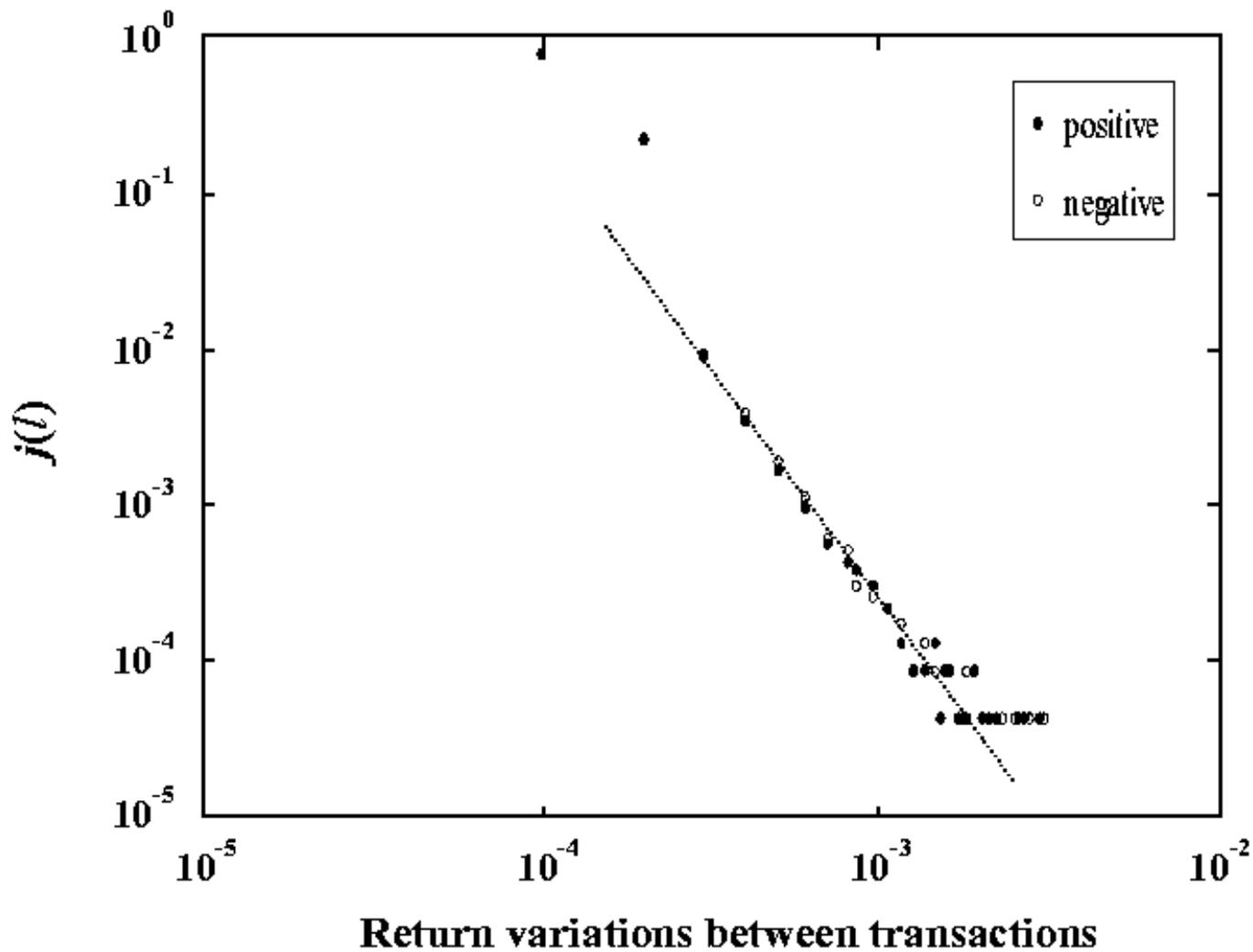

**Return variations between transactions**

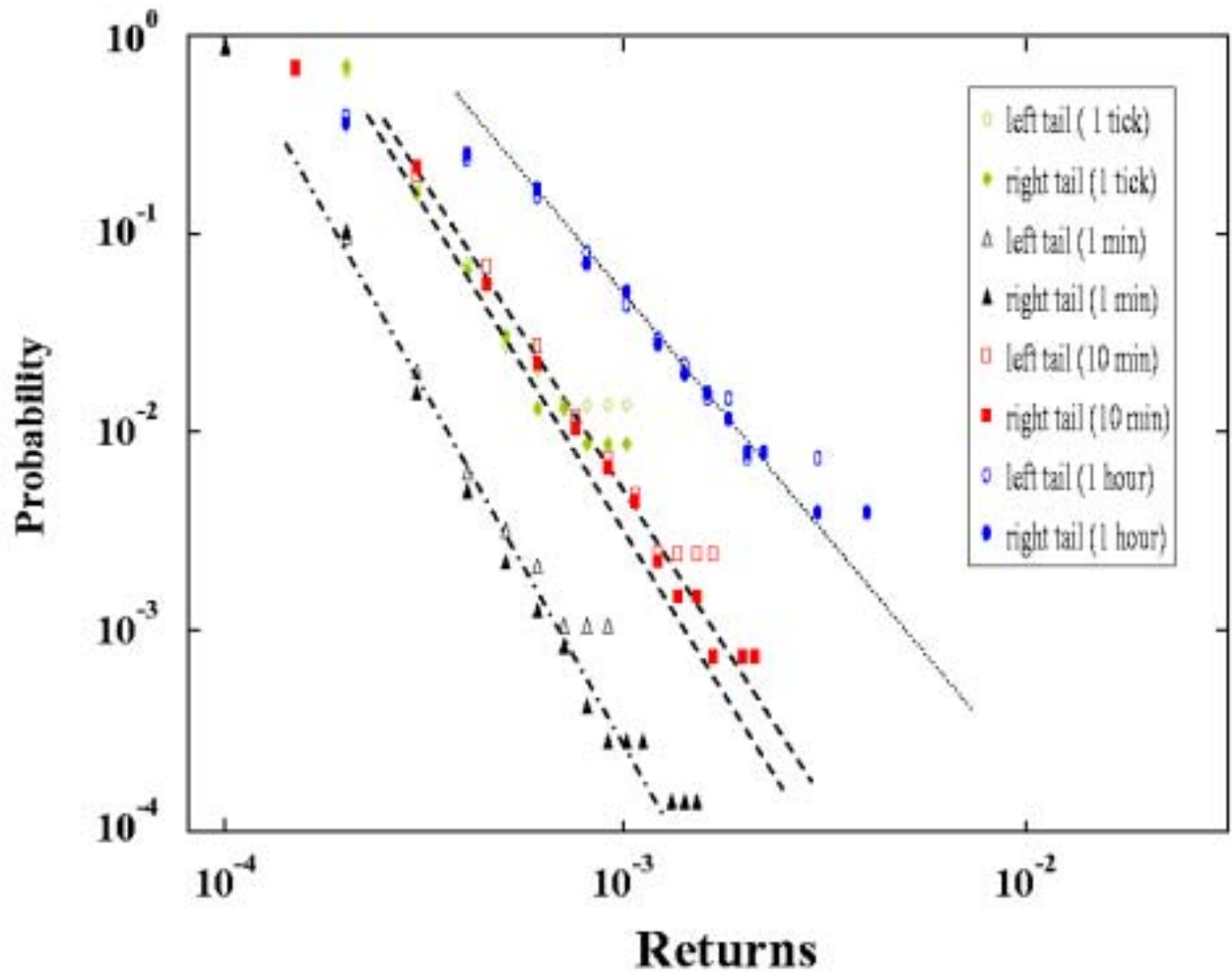

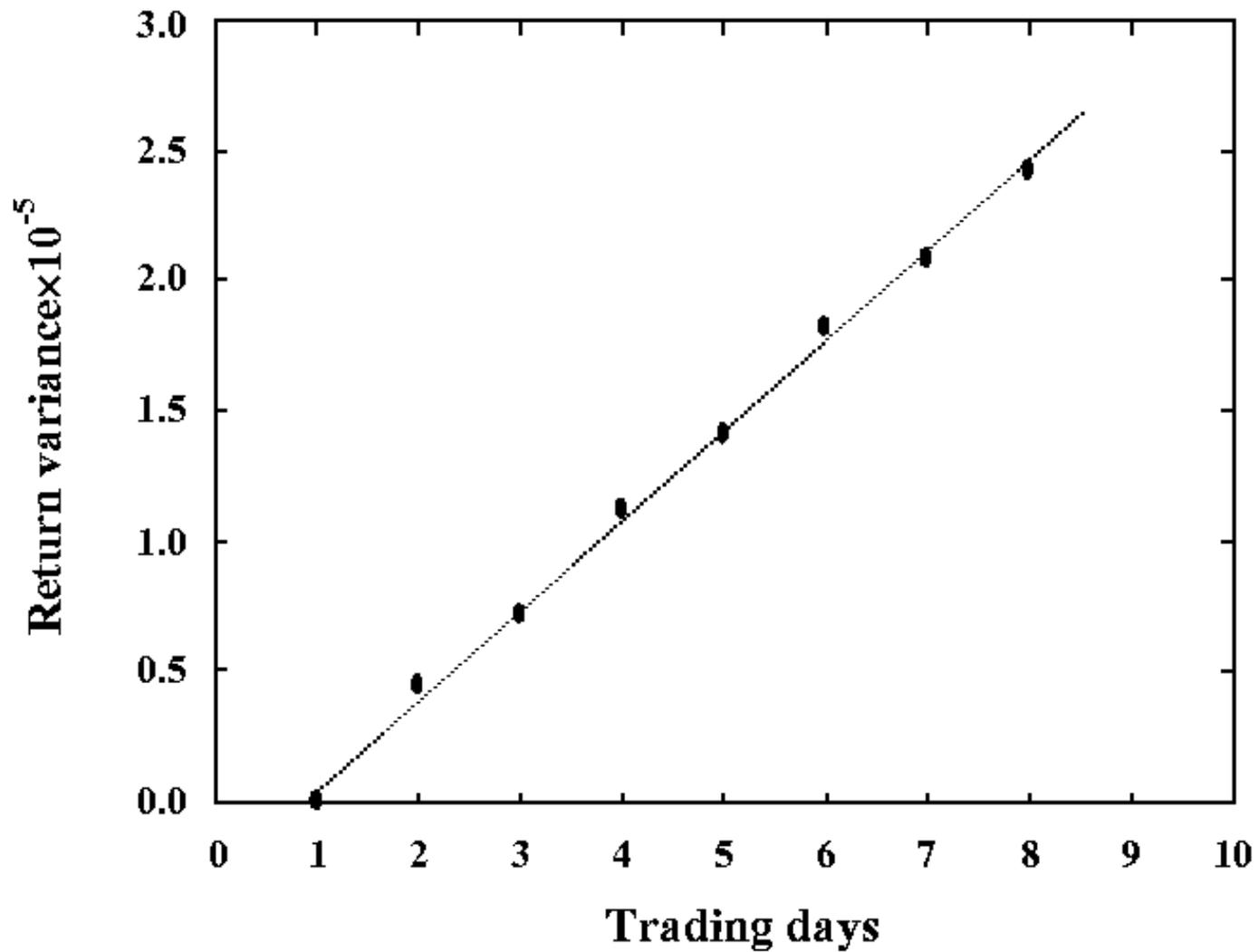